\documentclass[article]{aa}

\usepackage{natbib}
\usepackage{graphicx}
\usepackage{txfonts}

\begin{document}

\title{On the nature of kink MHD waves in magnetic flux tubes} 

\author{M. Goossens$^{1}$, J. Terradas$^{1}$, J.
Andries$^{1,2}$\thanks{Postdoctoral Fellow of the National Fund for
Scientific Research--Flanders (Belgium) (F.W.O.-Vlaanderen).}, I.
Arregui$^3$,
J. L. Ballester$^3$}

\offprints{M. Goossens, \email{Marcel.Goossens@wis.kuleuven.be}}
\institute{$^1$Centre Plasma Astrophysics and Leuven Mathematical Modeling and
Computational Science Center, Katholieke Universiteit Leuven, Leuven, B-3001,
Belgium, \email{Marcel.Goossens@wis.kuleuven.be}
\\ $^2$ Centre for Stellar and Planetary Astrophysics, Monash University, 3800
Victoria, Australia,
\email{Jesse.Andries@wis.kuleuven.be}
\\ $^3$ Departament de F\'\i sica,
Universitat de les Illes Balears, Spain, \email{Inigo.Arregui@uib.es}}
{}

\date{Received / Accepted }

\abstract{Magnetohydrodynamic (MHD) waves are often reported in the solar atmosphere and
usually classified as slow, fast, or Alfv\'en. The possibility that these waves have mixed
properties is often ignored.} {The goal of this work is to study and determine the nature
of MHD kink waves.} {This is done by calculating the frequency, the damping rate and the
eigenfunctions of MHD kink waves for three widely different MHD waves cases: a
compressible pressure-less plasma, an incompressible plasma and a compressible plasma with
non-zero plasma pressure which allows for MHD radiation.} {In all three cases the
frequency and the damping rate are for practical purposes the same as they differ at most
by terms proportional to $(k_z R)^2$. In the magnetic flux tube the kink waves are in all
three cases, to a high degree of accuracy incompressible waves with negligible pressure
perturbations and with mainly horizontal motions. The main restoring force of kink waves
in the magnetised flux tube is the magnetic tension force. The total pressure gradient
force cannot be neglected except when the frequency of the kink wave is equal or slightly
differs from the local Alfv\'{e}n frequency, i.e. in the resonant layer.}{Kink waves are
very robust and do not care about the details of the MHD wave environment. The adjective
fast is not the correct adjective to characterise kink waves. If an adjective is to be
used it should be Alfv\'{e}nic. However, it is better to realize that kink waves have
mixed properties and cannot be put in one single box.}


\titlerunning{Kink MHD waves in magnetic flux tubes}
\authorrunning{Goossens et al.}
\maketitle
\newcommand{\etal}{{et al.}}

\section{Introduction} The last decade has seen an avalanche of observations of
magnetohydrodynamic (MHD) waves in the solar atmosphere. It is clear now that
MHD waves are ubiquitous in the solar atmosphere. This has triggered new
theoretical research for explaining and interpreting the observed properties. A
special point of attention is whether these MHD waves are slow, fast or
Alfv\'{e}n waves. Apparently, a large fraction of the solar MHD waves community
favours very clear cut divisions and does not seem to appreciate the possibility
of MHD waves with mixed properties. The transverse oscillations observed in
coronal loops \citep[see for example][]{Aschwanden1999}, often triggered by a
nearby solar flare, are interpreted as fast kink MHD waves. A striking property
of these transverse waves is their fast damping with damping times of the order
of 3 - 5 periods. Resonant absorption is up to today the only damping mechanism
that offers a consistent explanation of this rapid damping.  Resonant absorption
relies on the transfer of energy from a global MHD wave to local resonant
Alfv\'{e}n waves. If this mechanism is indeed operational then this means that
the observed transverse oscillations have Alfv\'{e}nic properties in at least
part of the oscillating loop. The debate on the nature of MHD waves in the solar
atmosphere has gained new momentum when several groups, e.g.
\cite{DePontieu2007}, \cite{Okamoto2007}, \cite{Tomczyk2007} reported the
detection of Alfv\'{e}n waves in HINODE observations. Subsequently
\cite{VanDoorsselaere2008} argued strongly against the possible presence of
Alfv\'{e}n waves in the solar corona emphasising that Alfv\'{e}n waves cannot be
but torsional.  \cite{VanDoorsselaere2008}  compared fast kink MHD waves to
torsional Alfv\'{e}n waves and concluded that the HINODE observations have
nothing to do with Alfv\'{e}n waves but can be explained in terms of fast kink
MHD waves.

This paper will not try to explain the HINODE observations. Its aim is to
determine the nature of kink MHD waves on magnetic flux tubes. We have no doubt
about  the explanation of transverse oscillation of coronal loops in terms of
kink MHD waves. MHD waves with their azimuthal wave number equal to 1, i.e.
$m=1$, are the only motions that displace the axis of the loop and the loop as a
whole. It is not clear to us on what arguments the use of the adjective fast is
based. As far as we know there has not been any study of the forces that drive
the kink waves in coronal loops. If these waves are fast, then the pressure
gradient force should be, in general, the dominant force compared to the
magnetic tension force. We have to admit that we also have used the adjective
fast without a solid argument in favour of this classification. Our aim is to
understand the spatial structure of the motions in the kink waves. 

An MHD wave on an axi-symmetric 1-D cylindrical plasma equilibrium is
characterised by two wave numbers, the azimuthal wave number $m$, and the axial
wave number $k_z$. In addition modes can have a different number of nodes in the
radial direction and this number can be used to further classify the modes.
Hence, an MHD eigenmode is characterised by three numbers. The azimuthal wave
number is an integer. The modes with $m=0$ are usually called sausage (slow and
fast) or torsional (Alfv\'{e}n). The modes with $m=1$ are named kink and the
modes with $m \geq 2$ are flute modes. The axial wave number $k_z$  can be
discretised as $k_z = n \frac{\pi}{L}$ with $L$ the length of the loop and $n
=1,2, \ldots$ Depending on the dimensions of the equilibrium model there can be
more than one radial eigenmode for a given couple $(m, k_z)$. In what follows we
shall study linear MHD waves  that are superimposed on a flux tube in static
equilibrium with a straight and constant axial magnetic field. This equilibrium
model contains the essential physics of the problem and allows a relatively
straightforward mathematical analysis of the MHD waves. MHD waves have been
investigated in previous studies. However, these studies almost exclusively
focused on the frequencies of the MHD waves and in addition they were in most
cases restricted to real frequencies. For example, the paper by \cite{Edwin1983},
which is often referred to in the solar MHD wave community, is limited to real
frequencies and does not give any information on the eigenfunctions beyond the
fact that they can be expressed in terms of Bessel functions. Complex frequencies
were considered by \cite{Spruit1982} and by \cite{Cally1985}. \cite{Cally1985}
rightfully pointed out that essential physics is lost by restricting the analysis
to real frequencies. \cite{Spruit1982} is an exception to the rule in that he
discussed the eigenfunctions. In short, the eigenfunctions of MHD waves are not
well documented even for uniform equilibrium states and definitely not for 
non-uniform equilibrium states.

In order to illustrate the nature of MHD kink waves we shall study them for
three widely different MHD waves cases. The first case deals with compressible
MHD waves of a pressureless plasma on a high density flux tube embedded in a
low density magnetic plasma exterior. The assumption of a pressureless plasma
removes the slow waves from the analysis and the velocity and displacement have
no axial component. The density being higher in the flux tube than in the
exterior means that according to common wisdom the MHD waves are propagating
(body waves) in the flux tube and evanescent (surface wave) in the exterior.
The second case deals with  MHD waves of an incompressible plasma. The
assumption of incompressibility removes the fast waves. The waves are
evanescent (surface waves) both in the interior and exterior of the flux tube
and this behaviour is independent of the density being higher or lower in the
interior. In the third case we consider a situation where the compressible MHD
waves in a non-zero beta plasma are propagating in the  exterior. In that case
the MHD waves are damped by MHD radiation, but the kink MHD wave hardly feels
this wave leakage.

\section{Propagating kink MHD waves on dense pressureless flux tubes}

\subsection{Equations for MHD waves for  a pressureless plasma cylinder}
The equations for linear MHD waves on a 1-dimensional pressureless
cylinder with a straight field can be obtained from the more
general equations by e.g. \cite{Appert1974}; \cite{Sakurai1991a};
\cite{Goossens1992}, \cite{Goossens1995a} by putting the local
speed of sound $v_{\mathrm S}$ and the azimuthal component of the
equilibrium magnetic field $B_{\varphi}$ equal to 0. The resulting
equations are \citep[see e.g.][]{Goossens2008a}
\begin{eqnarray}
D\frac{ d(r \xi_r)}{ dr}&  = &  - C_2 r P',
\nonumber \\
\rho \left(\omega^2 - \omega_{\mathrm A}^2\right) \xi_r & = &
\frac{ d P'}{ dr},  \nonumber \\
\rho \left(\omega^2 - \omega_{\mathrm A}^2\right) \xi_{\varphi} & = & \frac{i m}{r} P',
\nonumber \\
\xi_{z} & = & 0, \nonumber \\
\nabla \cdot \vec{\xi} & = & \frac{ - P'}{
\rho v_{\mathrm A}^2}. \label{MHDwavesNoPressure1}
\end{eqnarray}
$\vec{\xi}$ is the Lagrangian displacement and $P'$ is the
Eulerian perturbation of total pressure. The coefficient functions
$D$ and $C_2$ in (\ref{MHDwavesNoPressure1}) are
\begin{eqnarray}
D & = & \rho  v_{\mathrm A}^2 \omega^2 \left(\omega^2 -
\omega_{\mathrm A}^2\right), \nonumber \\
C_2 & = & \omega^2 \left( \omega^2 - \omega_{\mathrm A}^2 -\frac{m^2}{r^2}v_{\mathrm A}^2\right).
\label{MHDwavesNoPressure2}
\end{eqnarray}
$\omega$ is the frequency and $m$ and $k_{z}$ are the azimuthal and
axial wave numbers, respectively. They define the dependence of the perturbed
quantities on time $t$ and on the ignorable spatial
variables ($\varphi, z$) as
\begin{equation}
\exp [i (m\varphi + k_{z}z - \omega t)] \label{t-phiz-dependency}.
\end{equation}
$\omega_{\mathrm A}$ and $v_{\mathrm A}$ are the local Alfv\'{e}n frequency and the
local Alfv\'{e}n velocity respectively. They are defined as
\begin{eqnarray}
\omega_{\mathrm A}^2 & = k_z^2 v_{\mathrm A}^2, \;\; v_{\mathrm A}^2 = \frac{ B^2}{ \mu \; \rho}, \;\;\omega_{\mathrm C}^2 =
\frac{ v_{\mathrm S}^2}{ v_{\mathrm S}^2 + v_{\mathrm A}^2 }\omega_{\mathrm A}^2,
\;\;  v_{\mathrm S}^2 = \frac{ \gamma p}{ \rho}.
\label{OmegaAC}
\end{eqnarray}
For completeness we have also listed the expression for the local cusp
frequency $\omega_{\mathrm C}$ and the local speed of sound $v_{\mathrm S}$. For a pressureless
plasma these two quantities vanish and as a consequence the slow modes are
removed from the system since $\xi_z = 0$.

The nature of an MHD wave is determined by the competition of the
restoring forces  which are the force due to the total (gas plus magnetic) pressure gradient and
the magnetic tension force. In ideal MHD we can obtain the
following expression for the Lorentz force, the magnetic tension
force $\vec{\Pi}$ and the magnetic pressure force $\nabla_\perp P'$ in
linear MHD waves on a background with a  constant  magnetic field
\begin{eqnarray}
\frac{ 1}{ \mu}  {(\nabla \times
\vec{B}')\times B\;\;\vec{1}_z} & = & - \left(\frac{d P'}{dr} + \rho
\omega_{\mathrm A}^2 \xi_{r}\right)\; \vec{1}_r\nonumber \\ &-& \left(i \frac{m}{r} P' + \rho
\omega_{\mathrm A}^2 \xi_{\varphi}\right) \; \vec{1}_{\varphi},
\nonumber \\
\vec{\Pi} & = & - \rho \omega_{\mathrm A}^2 \left(\xi_r \; \vec{1}_r +
\xi_{\varphi} \; \vec{1}_{\varphi} \right)= - \rho \omega_{\mathrm A}^2 \vec{\xi},
\nonumber \\
- \nabla_\perp P' & = & - \frac{ d P'}{
dr} \vec{1}_r  - \frac{i m}{r} P'
\vec{1}_{\varphi} \nonumber \\
& = & - \rho \left(\omega^2 - \omega_{\mathrm A}^2\right) \left(\xi_r \; \vec{1}_r + \xi_{\varphi}
\; \vec{1}_{\varphi}\right ) \nonumber \\
& = &
- \rho \left(\omega^2 - \omega_{\mathrm A}^2\right) \vec{\xi}.
\label{LF}
\end{eqnarray}
$\nabla_\perp $ is the gradient operator in horizontal planes
perpendicular to the constant vertical magnetic field. The last
equation in (\ref{LF}) is obtained by using the second and third
equations  of (\ref{MHDwavesNoPressure1})  to write   the
components of   $\nabla_\perp  P'$ in terms of the components of
$\xi$. Note also that in the third line of
(\ref{LF}) the first equality is general but the second equality
is not. A similar thing happens between the fourth and fifth line. 
The second equality in both cases  relies on the fact that
$\xi_z = 0$ which is the case for a pressureless plasma. The
important result from (\ref{LF}) is that the ratio of any of the
two relevant components (the radial or azimuthal) of pressure force to the corresponding
component of magnetic tension force is
\begin{equation}
\Lambda(\omega^2) = \frac{ \omega^2 - \omega_{\mathrm A}^2}
{ \omega_{\mathrm A}^2}. \label{RatioGradPrTf}
\end{equation}
In a non-uniform plasma with $v_{\mathrm A}^2$ and hence $\omega_{\mathrm A}^2$
dependent on position, $\Lambda$ also depends on position, meaning
that the nature of the MHD wave changes according to the properties
of the plasma it travels through. In a uniform plasma $\Lambda$ is a
constant and the nature of the wave does not change as it always
sees the same environment.

It is standard practice to rewrite the two first order ordinary
differential equations  of (\ref{MHDwavesNoPressure1}) as a
second order ordinary differential equation for $P'$:
\begin{equation}
\rho \left(\omega^2 - \omega_{\mathrm A}^2\right) \frac{ d }{
dr} \left \{\frac{ r} { \rho \left(\omega^2 -
\omega_{\mathrm A}^2\right)} \frac{ d P'} { dr} \right
\}  = \left \{ \frac{ m^2}{ r^2} -
\Gamma(\omega^2) \right \} r P', \label{PNoPr1}
\end{equation}
where $\Gamma(\omega^2)$ is the abbreviation for
\begin{equation}
\Gamma(\omega^2) =  \frac {\omega^2 - \omega_{\mathrm A}^2}{
v_{\mathrm A}^2}  = k_z^2 \;\;\Lambda(\omega^2).\label{Gamma}
\end{equation}
$\Gamma$ depends on $\omega^2$ but also on the equilibrium through
$v_{\mathrm A}^2$ and for a non-uniform plasma it is a function of position.
The sign of $\Gamma$, i.e. of $\Lambda$, determines the local
radial behaviour of the MHD wave. For  $\Gamma > 0$ the wave
behaves locally in the radial direction as a propagating wave, for
$\Gamma < 0$ the  the wave behaves locally as an evanescent wave (this is at
least valid at large distances from the tube).

\subsection{Pressureless flux tubes with uniform density}

For a uniform plasma we can rewrite (\ref{PNoPr1}) as
\begin{equation}
\frac{ d^2 P'}{ dr^2} +
\frac{ 1} { r} \frac{ d
P'}{ dr} - \left \{\frac{m^2}{r^2} - \Gamma(\omega^2)
\right \} P' = 0. \label{PNoPr2}
\end{equation}
Here $\Gamma(\omega^2)$ is constant. We now specialise on uniform
loops with a higher density $\rho_{\mathrm i}$ than the surrounding uniform
plasma which has a constant density $\rho_{\mathrm e} < \rho_{\mathrm i}$. At the loop
boundary $r= R$ the density $\rho$ changes discontinuously from
its internal value  $\rho_{\mathrm i}$ to its external value $\rho_{\mathrm e}$. Since
the magnetic field is uniform with the same strength everywhere,
it follows that
\begin{equation}
v_{\mathrm {Ai}}^2 = \frac{ B^2}{ \mu \rho_{\mathrm i}}
< v_{\mathrm {Ae}}^2 =
\frac{ B^2}{ \mu \rho_{\mathrm e}},\;\;\;\; \omega_{\mathrm {Ai}}^2 = k_z^2 v_{\mathrm {Ai}}^2  < \omega_{\mathrm {Ae}}^2 = k_z^2 v_{\mathrm {Ae}}^2.
\label{vAiVae}
\end{equation}
Note that the quantities $\Gamma(\omega^2)$ and
$\Lambda(\omega^2)$ are constant both in the interior and the
exterior of the flux tube but change discontinuously at the loop
boundary $r=R$. As our focus is on MHD waves that are propagating
in the interior of the loop and evanescent outside the loop, we
select the frequency so that
\begin{equation}
\Gamma_{\mathrm i}(\omega^2) > 0, \;\;\;\Gamma_{\mathrm e}(\omega^2) < 0,
\label{ProEvan1}
\end{equation}
which means that
\begin{equation}
\omega_{\mathrm {Ai}} ^2  < \omega^2 < \omega_{\mathrm {Ae}}^2.
\label{ProEvan2}
\end{equation}
This allows us to define radial wave numbers $k_{\mathrm i}$ and $k_{\mathrm e}$ as
\begin{equation}
k_{\mathrm i}^2 = \Gamma_{\mathrm i}(\omega^2) = \frac {\omega^2 - \omega_{\mathrm {Ai}}^2}
{ v_{\mathrm {Ai}}^2}, \;\;\; k_{\mathrm e}^2 = - \Gamma_{\mathrm e}(\omega^2) = -
\frac {\omega^2 - \omega_{\mathrm {Ae}}^2}{ v_{\mathrm {Ae}}^2}.
\label{kike}
\end{equation}
Equation~(\ref{PNoPr2}) can then be solved in terms of Bessel
functions $J_m(x)$  ($x = k_{\mathrm i} r$) in the internal part of the flux
tube and $K_m(y)$ ($ y = k_{\mathrm e} r$) in the exterior region.
\begin{eqnarray}
 P'_{\mathrm i}(r) & = & \alpha J_m(x), \nonumber \\
\xi_{r, \mathrm{i}}(r) & = & \alpha \frac{ k_{\mathrm i}}{ \rho
(\omega^2 - \omega_{\mathrm {Ai}}^2)}
\;J'_m(x), \nonumber \\
 P'_{\mathrm e}(r) & = & \beta K_m(y), \nonumber \\
\xi_{r, \mathrm{e}}(r) & = & \beta \frac{ k_{\mathrm e}} { \rho
(\omega^2 - \omega_{\mathrm {Ae}}^2)} \;K'_m(y). \label{SolutionsNoPr1}
\end{eqnarray}
The prime denotes a derivative with respect to the argument $x$ or
$y$ and  $\alpha$ and $\beta$ are constants. Continuity of total pressure and
the radial component of the Lagrangian displacement leads to the
dispersion relation:
\begin{equation}
 F \; \frac{ J'_m(x_0) K_m(y_0)}{
 J_m(x_0) K'_m(y_0)} = 1 \label{DRNoPr1},
\end{equation}
with the quantity $F$ given by
\begin{equation}
F = \frac{ k_{\mathrm i}}{ k_{\mathrm e}}
\frac{ \rho_{\mathrm e} (\omega^2 - \omega_{\mathrm {Ae}}^2)}
{ \rho_{\mathrm i} (\omega^2 - \omega_{\mathrm {Ai}}^2)}, \label{F}
\end{equation}
being $x_0 = k_{\mathrm i} R$ and $y_0 = k_{\mathrm e} R$. The dispersion relation
(\ref{DRNoPr1}) can be solved numerically. This was done for real
frequencies by \cite{Edwin1983} and for complex frequencies by
e.g. \cite{Spruit1982} and \cite{Cally1985, Cally2003}. However, it is
instructive and also accurate to consider the so-called thin tube
(TT) approximation ($k_zR\ll1$). The Bessel functions $J_m(x)$ and $K_m(y)$  in
(\ref{DRNoPr1}) are replaced with their first order asymptotic
expansions. The dispersion relation (\ref{DRNoPr1}) is reduced to
\begin{eqnarray}
1 + F \;\frac{ k_{\mathrm e}}{ k_{\mathrm i}} = 0.
\label{DRNoPr2}
\end{eqnarray}
The solution for the frequency is
\begin{equation}
\omega^2  =  \frac{ \rho_{\mathrm i}  \omega_{\mathrm {Ai}}^2 + \rho_{\mathrm e}
\omega_{\mathrm {Ae}}^2} { \rho_{\mathrm i} + \rho_{\mathrm e}} = \omega_{\mathrm k}^2,
\label{Freqkink}
\end{equation}
and for the radial wave numbers $k_{\mathrm i}$ and $k_{\mathrm e}$
\begin{equation}
k_{\mathrm i}^2  =  k_{\mathrm e}^2 = k_z^2 \frac{ \rho_{\mathrm i} - \rho_{\mathrm e}}
{ \rho_{\mathrm i} + \rho_{\mathrm e}}. \label{kekiNoPr}
\end{equation}
The right hand side of (\ref{Freqkink}) is almost invariably called
the square of the kink frequency and denoted as $\omega_{\mathrm k}^2$. In the
thin tube approximation the frequency is independent of the wave
number $m\geq 1$ as already noted \cite{Goossens1992}. Hence all
flute modes with $m\geq 2$ have the same frequency as the kink mode
with $m=1$. The radial wave numbers $k_{\mathrm i}$ and $k_{\mathrm e}$ depend in a
simple way on the density contrast. As the density contrast
decreases, radial propagation is increasingly impeded and the nature
of the MHD wave becomes gradually more Alfv\'{e}nic, for $\rho_{\mathrm i} =
\rho_{\mathrm e}$ it is purely Alfv\'{e}nic as the frequency of the global
wave is equal to the local Alfv\'{e}n frequency everywhere. This is
confirmed by the value of $\Lambda$
\begin{equation}
\Lambda_{\mathrm i} (\omega^2) = - \Lambda_{\mathrm e} (\omega^2) =
\frac{ \rho_{\mathrm i} - \rho_{\mathrm e}} { \rho_{\mathrm i} +
\rho_{\mathrm e}}. \label{RatioGradPrT2}
\end{equation}
$\Lambda(\omega^2)$ is constant in both the interior and exterior. It is
positive in the interior meaning that the magnetic pressure force and the
magnetic tension force act in the same direction. In the exterior its value is
the exact opposite of that in the interior. The magnetic pressure force and the
magnetic tension force now oppose each other. Equation~(\ref{RatioGradPrT2}) shows that
the MHD waves are always dominated by magnetic tension forces and that they are
predominantly Alfv\'{e}nic in nature. Take as an example a density contrast
$\rho_{\mathrm i}/\rho_{\mathrm e} = 3$ then $\Lambda_{\mathrm i} = 1/2$, $\Lambda_{\mathrm e} = -1/2$ so that the
magnetic tension force is always twice as important as the pressure force.

The TT approximations to the eigenfunctions are
\begin{eqnarray}
\frac{ \xi_{r, \mathrm{i}}(r)}{ R} & = & C, \nonumber \\
\frac{ \xi_{\varphi, \mathrm{i}}(r)}{ R} & = & i \;C,
\nonumber \\
\frac{ P'_{\mathrm i}(r)}{ (B^2/\mu)} & = & C
\;(k_z R)^2 \; \frac{
\rho_{\mathrm i} - \rho_{\mathrm e}}{ \rho_{\mathrm i} + \rho_{\mathrm e}} \;\frac{ r}
{ R},\nonumber \\
\nabla \cdot \vec{\xi}_{\mathrm i} & = & - C \;(k_z R)^2 \;\frac{
\rho_{\mathrm i} - \rho_{\mathrm e}}
{ \rho_{\mathrm i} + \rho_{\mathrm e}} \; \frac{ r}{ R},
\nonumber \\
\xi_{\varphi, \mathrm{e}}(R_{>}) & = & - \xi_{\varphi, \mathrm{i}}(R_{<}).
\label{SolutionsNoPr2}
\end{eqnarray}
Note that when deriving (\ref{SolutionsNoPr2}) we have omitted terms
of order $(k_z R)^2$ and higher unless the terms of order $(k_z
R)^2$ are the first non-vanishing contribution to the expression
under study. For example the expressions for $\xi_{r, \mathrm{i}}(r)/ R$ and
$\xi_{\varphi, \mathrm{i}}(r)/ R$ mean that these two quantities are equal up
to differences of order $(k_z R)^2$. The eigenfunctions are
determined up to a multiplicative constant $C$ which can be used to
specify e.g. the radial displacement of the boundary of the loop.
The radial and azimuthal components are $\pi/2$ out of phase but
they have equal magnitudes and in the loop they are constant. The
wave is in the propagating domain in the internal part of the loop
($k_{\mathrm i}^2 >0$)  but there are not any spatial variations.  For
realistic values of $k_z R = (\pi R)/L $ the pressure perturbation
and the divergence of the displacement field are zero for all practical
purposes. The kink mode is to a high degree of accuracy an
incompressible wave with very small magnetic pressure perturbations. Apart
from $\xi_{\varphi}$ the wave quantities are continuous at $r= R$.
$\xi_{\varphi}$ varies in a discontinuous manner at $r=R$ with
opposite values at $R_<$ and $R_>$. This discontinuous behaviour is
due to the change of sign of the factor $\omega^2 - \omega_{\mathrm A}^2$ when
we move from the interior to the exterior of the loop. The behaviour of $\xi_{\varphi}$ creates strong shear
layers which might undergo Kelvin-Helmholtz type instabilities as
shown by \cite{Terradas2008}. 

So far we have described the properties of modes that involve non-zero total
pressure ($P'\neq 0$). However, if the medium is uniform the system of equations
given by (\ref{MHDwavesNoPressure1}) also allows pure incompressible Alfv\'en
waves. They are only driven by magnetic tension, their eigenfrequency is simply
$\omega=\omega_{\mathrm A}$ and the total pressure, $P'$, is equal to zero. To have such
modes the displacement has to satisfy $\nabla \cdot \vec{\xi} =0$. In
cylindrical coordinates this means that,  \begin{equation} \frac{d
\left(r\xi_r\right)}{dr}+i m \xi_\varphi=0. \label{Alfincompress} \end{equation}
If we prescribe the radial dependence of one of the components of the
displacement, we can easily calculate the other component from the previous
equation. The case $m=0$ is a particular solution that represents torsional
Alfv\'en waves ($\xi_r=0$, $\xi_\varphi$ arbitrary), but equation
(\ref{Alfincompress}) can be solved for any azimuthal wavenumber $m$. Let us
concentrate on $m=1$ and the homogeneous loop model. Now we can have two
different incompressible Alfv\'en waves. One inside oscillating at the
frequency  $\omega_{\mathrm {Ai}}$ and another outside oscillating at $\omega_{\mathrm {Ae}}$.
However, it is important to note that in  such a configuration the radial
displacement of the internal and external Alfv\'en waves has to vanish at $r=R$
(otherwise the continuity of the radial component is not guaranteed), i.e. the
modes are localised in regions of constant Alfv\'en frequency. This means that
an internal incompressible Alfv\'en wave is unable to laterally displace the
full tube (it cannot displace the tube boundary), although it is able to produce
an incompressible motion of the loop axis at its surroundings (for $m=1$). Since
these pure incompressible Alfv\'en waves do not move the whole tube hereafter we
will focus again on the kink solutions with $P'\neq 0$. Contrary to the
incompressible Alfv\'en waves these waves (with $P'\neq 0$) are able to connect
the internal and external medium and to produce a coherent motion of the system
because of their mixed nature. The fact that $P'$ is small but different from
zero plays a fundamental role in the mixed properties of these kink waves.


\subsection{Beyond the TT approximation for pressureless uniform flux tubes}

The analytic expressions (\ref{SolutionsNoPr2}) ($P'\neq 0$) have been obtained
in the limit $k_z R << 1$. It is straightforward to solve the dispersion relation
(\ref{DRNoPr1}) and calculate the spatial solutions (\ref{SolutionsNoPr1}). This
allows us to determine how the analytical expressions are modified by effects due
to a finite radius. In Figure \ref{eigen} the eigenfunctions of three loops with
different radii are represented. It is clear that the spatial profile is well
described by the approximated solutions in the TT limit given by
equations~(\ref{SolutionsNoPr2}). The radial and azimuthal components are
constant inside the loop, the azimuthal component has the expected jump at $r=R$,
while the total pressure grows linearly with the radius. Increasing  $R$ results
in an increase of the total pressure, and thus compressibility, since this
magnitude is proportional to $(k_zR)^2$. Interestingly, for fat loops (see the
case $R/L=0.1$) the TT approximations of the eigenfunctions are still quite
valid. An analysis of the forces (not shown here) indicates that, even for thick
loops, the tension dominates over the magnetic pressure gradient. 

\begin{figure}[hh] \center{
\resizebox{7.cm}{!}{\includegraphics{f1a.eps}}
\resizebox{7.cm}{!}{\includegraphics{f1b.eps}}
\resizebox{7.cm}{!}{\includegraphics{f1c.eps}}
} 

\caption{ \small Radial dependence of the normalised eigenfunctions, $\xi_{r}$
(real),
$\xi_{\varphi}$ (imaginary), and $P'$ (real) calculated using~(\ref{SolutionsNoPr1}) after solving the
dispersion relation~(\ref{DRNoPr1}). The different lines represent  three different loop
radii, $R/L=0.01,0.02,0.1$. The corresponding dimensionless eigenvalues are $\omega
L/v_{\mathrm {Ai}}=3.847,3.844, 3.803$. The circles describe the incom\-pressible eigenfunctions
calculated using~(\ref{SolutionsInC1}) after solving the dispersion relation given
by~(\ref{DRInC1}). The frequencies for the three loop radii are, in the incompressible
limit, $\omega L/v_{\mathrm {Ai}}=3.846,3.842, 3.785$. In these calculations $\rho_{\mathrm i}/\rho_{\mathrm e} = 3$ and
$k_z=\pi/L$.} \label{eigen} \end{figure}

\subsection{Pressureless flux tubes with non-uniform density} 

In this subsection we remove the discontinuous variation of density from its
internal value $\rho_{\mathrm i}$ to $\rho_{\mathrm e}$ by a continuous variation in a non-uniform
layer $[R - l/2 , R + l/2]$. A fully non-uniform equilibrium state corresponds to
$l= 2 R$.  When the jump in $\omega_{\mathrm A}$ is replaced by a continuous variation of
$\omega_{\mathrm A}$ new physics is introduced in the system. The continuous variation of
$\omega_{\mathrm A}$ has the important effect that the kink MHD wave, which has its
frequency in the Alfv\'{e}n continuum, interacts with local Alfv\'{e}n continuum
waves and  gets damped. This resonant damping is translated in a complex
frequency (and complex eigenfunction). In the present paper damped global
eigenmodes that are coupled to resonant Alfv\'{e}n waves in a non-uniform
equilibrium state shall be computed by two methods. The first method is to use a
numerical code that integrates the resistive MHD equations in the whole volume of
the equilibrium state to determine a selected mode or part of the resistive
spectrum of the system \citep[see for example][]{VanDoorsselaere2004,
Arregui2005,Terradas2006}. The second method was introduced by \cite{Tirry1996}.
It circumvents the numerical integration of the non-ideal MHD equations and only
requires numerical integration (or closed analytical solutions) of the linear
ideal MHD equations. The method relies on the fact  that dissipation is important
only in a narrow layer around the resonant point where the real part of the
frequency of quasi-mode equals the local Alfv\'{e}n frequency. This makes it
possible to obtain analytical solutions to simplified versions of the linear
dissipative MHD equations which accurately describe the linear motions in the
dissipative layer and in two overlap regions to the left and right of the
dissipative layer. In these overlap regions both ideal MHD and dissipative MHD
are valid. Asymptotic analysis of the analytical dissipative solutions allows to
derive jump conditions that can be used to connect the solutions to the left and
right of the dissipative layer. The jump conditions were derived by e.g.
\cite{Sakurai1991a}, \cite{Goossens1992}, \cite{Goossens1995a} and
\cite{Goossens1995b} for the driven problem and by \cite{Tirry1996} for the
eigenvalue problem. A schematic overview of the various regions involved in this
method is shown in Figure 1 of \cite{Stenuit1998}. This method was used for
computing eigenmodes of various non-uniform plasma configurations by e.g.
\cite{Tirry1998a,Tirry1998b}, \cite{Stenuit1998,Stenuit1999},
\cite{Andries2000,Andries2001}. A related method was used by \cite{Sakurai1991b}
and \cite{Stenuit1995} for the computation of resonant Alfv\'{e}n waves in the
driven problem with a prescribed and real frequency and by \cite{Keppens1994}
for  computing the multiple scattering and resonant absorption of p-modes by 
fibril sunspots. Comparison with results of fully dissipative computations  show
that the method is very accurate.

A drastic variant of the method that avoids solving the non-ideal
MHD equations uses the so-called thin boundary (TB) approximation.
In this lazy version the ideal MHD equations are not solved in the
non-uniform plasma but the plasma is treated as if it were uniform
all the way up to the dissipative layer. This is definitely a very
strong assumption since the thickness of the dissipative layer is
measured by the quantity $\delta_{\mathrm A}$:
\begin{equation}
 \delta_{\mathrm A} = \left ( \frac{ \omega \eta}{
 \mid \Delta_{\mathrm A} \mid }
 \right )^{1/3},\;\;  \Delta_{\mathrm A} = \frac{d}{dr}\left(\omega^2 - \omega_{\mathrm A}^2\right).
 \label{delta}
\end{equation}
Here $\eta$ is the magnetic resistivity. If we denote $l$ the typical length scale for the variations of the
equilibrium quantities then
\begin{equation}
\frac{ l}{ \delta_{\mathrm A}}  = (R_m)^{1/3},
\label{ldelta}
\end{equation}
where $R_m$ is the magnetic Reynolds number ($R_m \sim \eta^{-1}$). Strictly speaking the
TB approximation assumes that the nonuniform layer and the
dissipative layer coincide. It is common practice not to draw
attention to this assumption but instead to refer to the TB
approximation as the approximation that adopts $l/R << 1$. This TB
approximation was first used by \cite{Hollweg1988}. A discussion of
the TB approximation can be found in \cite{Goossens2008a}.

In the TB approximation we need to add an additional term to the
dispersion relation which takes into account the jump in the radial
component across the resonant layer where the real part of the kink
eigenmode is equal to the local Alfv\'{e}n frequency $\omega =
\omega_{\mathrm A}(r_{\mathrm A})$. $r_{\mathrm A}$ is the resonant position which in the thin
boundary approximation $r_{\mathrm A} = R$. The jump in $\xi_r$ is
\begin{equation}
[\xi_r] = -i \pi \frac{ m^2 / r_{\mathrm A}^2}{ \rho
\mid \Delta_{\mathrm A} \mid } P', \;\; [P']= 0.  \label{JumpNoPr}
\end{equation}
The modified version of the ideal dispersion relation (\ref{DRNoPr1})
is
\begin{equation}
F \;\frac{ J'_m(x_0) K_m(y_0)}{ J_m(x_0)
K'_m(y_0)} -i \; G \; \frac{ K_m(y_0)}{
K'_m(y_0)}
= 1.
\label{DRNoPr3}
\end{equation}
$F$ is given by (\ref{F}) and $G$ is defined as
\begin{equation}
G = \pi \frac{ m^2 / r_{\mathrm A}^2}{\rho \mid \Delta_{\mathrm A} \mid }
\frac{ \rho_{\mathrm e} \left(\omega^2 -
\omega_{\mathrm {Ae}}^2\right)}{ k_{\mathrm e}}. \label{G}
\end{equation}
$G$ contains the effect of the resonance. When we combine the thin
tube (TT) approximation with the thin boundary (TB) approximation,
we can simplify the dispersion relation to
\begin{equation}
1 + F \; \frac{ k_{\mathrm e}}{ k_{\mathrm i}} - i \;G \;
\frac{ k_{\mathrm e} R}{ m} = 0. \label{DRNoPr4}
\end{equation}
The zero order solution to (\ref{DRNoPr4}), i.e. the solution when
the effect of the resonance is not taken into account is of course
(\ref{Freqkink}). In order to take the effect of the resonance into
account we write
\begin{equation}
\omega = \omega_{\mathrm R} + i \gamma, \;\;\omega_{\mathrm R} = \omega_{\mathrm k},
\label{OmegaComplex}
\end{equation}
and approximate $\omega^2$ with $\omega_{\mathrm k} ^2 + 2 i \omega_{\mathrm k} \gamma$.
The solution for the damping decrement is
\begin{equation}
\frac{ \gamma}{  \omega_{\mathrm k}} =
-\frac{ \pi/2}{ \omega_{\mathrm k}^2}
\frac{ m}{ R} \frac{
\rho_{\mathrm i}^2 \rho_{\mathrm e}^2}{ (\rho_{\mathrm i} + \rho_{\mathrm e})^3}
\frac{ (\omega_{\mathrm {Ai}}^2 - \omega_{\mathrm {Ae}}^2)^2}{\rho(r_{\mathrm A})
\mid \Delta_{\mathrm A}(r_{\mathrm A}) \mid}. \label{GammaNoPr1}
\end{equation}
Equation~(\ref{GammaNoPr1}) agrees with equation~(77) of
\cite{Goossens1992} when that equation is corrected  for a typo as
the factor $\left(\omega_{\mathrm {Ai}}^2 - \omega_{\mathrm {Ae}}^2\right)$ should be squared.
This is surprising since that result was obtained by
\cite{Goossens1992} for surface waves in incompressible plasmas. In
the same section of that paper it was noted that  there is no
distinction between  compressible and incompressible plasmas for
surface waves on thin tubes.

Equation~(\ref{GammaNoPr1}) shows that the damping decrement depends
linearly on $m$. Since we are mainly interested in $m=1$ we shall
specialise to that value in the remainder of this subsection. If the
variation of $\omega_{\mathrm A}^2$ is solely due to the variation of density
$\rho$ as is the case here since we have considered a constant
vertical magnetic field, equation~(\ref{GammaNoPr1}) can be
rewritten as
\begin{equation}
\frac{ \gamma}{  \omega_{\mathrm k}} = -
\frac{ \pi} { 8} \frac{
m}{ R} \frac{ (\rho_{\mathrm i} - \rho_{\mathrm e})^2}
{ \rho_{\mathrm i} + \rho_{\mathrm e} } \frac{ 1}{ \mid
\frac{ d \rho}{ dr} \mid_{r_{\mathrm A}}}.
\label{GammaNoPr2}
\end{equation}
For a linear profile of density
\begin{eqnarray*}
\mid \frac{ d \rho}{ dr}\mid_{r_{\mathrm A}} =
\frac{ \rho_{\mathrm i} - \rho_{\mathrm e}}{ l},
\end{eqnarray*}
so that
\begin{equation}
\frac{ \gamma}{  \omega_{\mathrm k}}  = -
\frac{ \pi}{ 8} \frac{
l}{ R}\frac{ \rho_{\mathrm i} -
\rho_{\mathrm e}}{ \rho_{\mathrm i} + \rho_{\mathrm e}}, \;\;\; \frac{
\tau_{\mathrm D}}{ T} = \frac{ 4}{
\pi^2} \frac{ 1}{ l/R}
\frac{ \rho_{\mathrm i} + \rho_{\mathrm e}}{ \rho_{\mathrm i} -
\rho_{\mathrm e}}.
 \label{GammaNoPr3}
\end{equation}
In (\ref{GammaNoPr3}) $\tau_{\mathrm D}$ is the damping time and $T$ the
period. Note that the result  for $\frac{
\gamma}{  \omega_{\mathrm k}}$  of equation~(\ref{GammaNoPr3}) agrees with
equation~(79b) of \cite{Goossens1992}.

For a sinusoidal profile of density
\begin{eqnarray*}
\mid \frac{ d \rho}{ dr}\mid_{r_{\mathrm A}} =
\frac{ \pi}{ 2} \frac{
\rho_{\mathrm i} - \rho_{\mathrm e}}{ l},
\end{eqnarray*}
so that
\begin{eqnarray}
\frac{ \gamma}{  \omega_{\mathrm k}} = -
\frac{ 1} { 4} \frac{
l}{ R}\frac{ \rho_{\mathrm i} -
\rho_{\mathrm e}}{ \rho_{\mathrm i} + \rho_{\mathrm e}}, \;\;\; \frac{
\tau_{\mathrm D}}{ T} & = & \frac{
2}{ \pi} \frac{ 1}{ l/R}
\frac{ \rho_{\mathrm i} + \rho_{\mathrm e}}{ \rho_{\mathrm i} -
\rho_{\mathrm e}}.
 \label{GammaNoPr4}
\end{eqnarray}
Here the results agree with those obtained by \cite{Ruderman2002}. 
At this point we like to stress that the TTTB approximation turns
out to be remarkably accurate far beyond its domain of
applicability. This is clearly illustrated in a recent analytical
seismological study by \cite{Goossens2008b} which complemented a
fully numerical seismology investigation by \cite{Arregui2007}.

The eigenfunctions in the thin dissipative layer can be described by
the functions $F(\tau)$ and $G(\tau)$ defined by
\cite{Goossens1995a} for the driven problem and the functions
$\tilde{F}(\tau)$ and $\tilde{G}(\tau)$ defined by \cite{Ruderman1995}
for the incompressible eigenvalue problem and by \cite{Tirry1996}
for the compressible eigenvalue problem. In the dissipative layer the
MHD kink waves are highly Alfv\'{e}nic. This can be understood as follows.
From the analysis by \cite{Sakurai1991a}, \cite{Goossens1995a} and
\cite{Tirry1996} it follows that in the dissipative layer, the
Eulerian perturbation of total pressure $P'$ is constant and that
$\mid \xi_r \mid << \mid \xi_{\varphi} \mid $. We do not have to
worry about $\xi_z$ since it is zero for a pressureless plasma. In
addition
\begin{equation}
\xi_{\varphi}(\tau) = \frac{ m \; P'}{ r\;
\rho} \frac{ \tilde{F}(\tau)}{\delta_{\mathrm A} \mid \Delta_{\mathrm A}
\mid }.  \label{xiDL}
\end{equation}
Here $\tau = \left(r - r_{\mathrm A}\right)/ \delta_{\mathrm A}$ is the stretched independent
variable used in the dissipative layer which has a typical width
$[-5\delta_{\mathrm A},  5\delta_{\mathrm A}]$. $\tilde{F}(\tau)$ is a complex function
and $\xi_{\varphi}(\tau)$ does not suffer a discontinuous jump as in
the case of a uniform plasma. It is characterised by rapid spatial
variation in the dissipative layer. From (\ref{xiDL}) it follows
that the ratio of the $\varphi$ component of magnetic tension to
that of pressure gradient is
\begin{equation}
\frac{ - \rho \; \omega^2_{\mathrm A} \; \xi_{\varphi}}{- i
(m/r_{\mathrm A})  P'} = - i \frac{ \omega^2_{\mathrm A}}{\mid \Delta_{\mathrm A}
\mid } \frac{ \tilde{F}(\tau)}{ \delta_{\mathrm A} }
\approx - i \; \frac{ l}{ \delta_{\mathrm A}} \;
\tilde{F}(\tau), \label{MTMPr1}
\end{equation}
where we have used $ \omega^2_{\mathrm A} /\mid \Delta_{\mathrm A} \mid   \approx l$.
Since  $\tilde{F}(\tau)$ is of order unity or bigger it follows that
\begin{equation}
\mid \frac{ - \rho \; \omega^2_{\mathrm A} \; \xi_{\varphi}}{- i
(m/r_{\mathrm A})  P'} \mid \approx (R_m)^{1/3} >> 1. \label{MTMPr2}
\end{equation}
In the dissipative layer the magnetic tension force is far bigger than the
pressure gradient force.  Hence the MHD kink wave is highly Alfv\'{e}nic in the
dissipative layer.

\subsection{Beyond the TTTB approximation for pressureless non-uniform flux
tubes} 

In this subsection  we go beyond the thin boundary approximation and we consider
thick non-uniform layers with $l/R = 0.2$ and $0.4$. The eigenfunctions are
numerically calculated by solving the full set of linear, resistive MHD equations
described in \citet{Terradas2006} and using the PDE2D code \citep{Sewell2005}.
Figure~\ref{eigennonunif} displays the obtained results. It is clear that the two
inhomogeneous solutions are almost identical to the homogeneous solution, except
in the non-uniform layer, where large displacements are found. This is the
location where the resonance takes place.

In Figure~\ref{forces} we see that both the radial and azimuthal components of
the Lorentz force are dominated by magnetic tension. The magnitude of the
magnetic tension and pressure is of the same order (with the tension about twice
as important as the magnetic pressure force such as corresponds to a density
contrast of 3) except in the inhomogeneous layer
where the tension in the azimuthal direction is clearly dominant, reflecting the
strong Alfv\'enic nature of the solution in the dissipative layer.

\begin{figure}[hh] \center{
\resizebox{7.cm}{!}{\includegraphics{f2a.eps}}
\resizebox{7.cm}{!}{\includegraphics{f2b.eps}}
\resizebox{7.cm}{!}{\includegraphics{f2c.eps}}
}

\caption{ \small Radial dependence of the moduli of the eigenfunctions, $\xi_{r}$,
$\xi_{\varphi}$, and $P'$ calculated numerically. The continuous lines represent the
a\-na\-lytical solutions for $l=0$ computed with~(\ref{SolutionsNoPr1}) (see also
Figure~\ref{eigen}). Dashed and dotted lines represent the numerical computed
eigenfunctions ($R_m=10^6$) for $l/R=0.2$ and $l/R=0.4$, respectively. The corres\-ponding
dimensionless eigenvalues are $\omega_{\mathrm R} L/v_{\mathrm {Ai}}=3.847,3.799,3.795$ and $\omega_{\mathrm i}
L/v_{\mathrm {Ai}}= 0,0.092,0.192$. In these calculations $\rho_{\mathrm i}/\rho_{\mathrm e} = 3$, $k_z=\pi/L$ and
$R/L=0.1$.}

\label{eigennonunif} \end{figure}

\begin{figure}[hh] \center{
\resizebox{7.cm}{!}{\includegraphics{f3a.eps}}
\resizebox{7.cm}{!}{\includegraphics{f3b.eps}}
\resizebox{7.cm}{!}{\includegraphics{f3c.eps}}
} 

\caption{ \small {\em Top panel:} Moduli of the radial forces. {\em Middle
panel:} Moduli of the azimuthal forces.  {\em Bottom panel:} Ratio of the
modulus of the magnetic tension to the magnetic pressure gradient in the radial and the
a\-zi\-muthal direction. This case corresponds to the eigensolution displayed in
Figure~\ref{eigennonunif} with $l/R=0.4$. The vertical dashed lines show the limits of the
non-uniform layer. \label{forces}}
\end{figure}

\section{Incompressible MHD waves on flux tubes}

\subsection{Equations for incompressible MHD waves on a plasma cylinder}

The equations for incompressible MHD waves can be found in
\cite{Goossens1992}. Incompressibility means that we take the
limit $v_{\mathrm S} \rightarrow \infty$ and enforce $\nabla \cdot \vec{\xi}
= 0.$ Note also that in the incompressible case $\omega_{\mathrm C} =
\omega_{\mathrm A}$.  The relevant equations are

\begin{eqnarray}
\rho \left(\omega^2 - \omega_{\mathrm A}^2\right)\frac{ d(r \xi_r)}{ dr}
&  = & \left(\frac{m^2}{r^2} + k_z^2\right) r P',
\nonumber \\
\rho \left(\omega^2 - \omega_{\mathrm A}^2\right) \xi_r  & = &
\frac{ d P'}{ dr}, \nonumber \\
\rho \left(\omega^2 - \omega_{\mathrm A}^2\right) \xi_{\varphi} & = & \frac{i m}{r} P',
\nonumber \\
\rho \left(\omega^2 - \omega_{\mathrm A}^2\right) \xi_{z} & = & i k_{z} P'.
\label{MHDwavesInCom1}
\end{eqnarray}
As before we concentrate on solutions with $P'\neq 0$ (although now we assume
incompressibility) and rewrite (\ref{MHDwavesInCom1}) as a second order
ordinary differential equation for $P'$ (eliminating the displacements):
\begin{equation}
\rho \left(\omega^2 - \omega_{\mathrm A}^2\right) \frac{ d }{
dr} \left \{\frac{ r}{ \rho \left(\omega^2 -
\omega_{\mathrm A}^2\right)} \frac{ d P'}{ dr} \right \}
=
 \left \{ \frac{m^2}{r^2}+ k_z^2 \right \} r P'.
\label{PInC1}
\end{equation}
Here we do not need to worry about propagating and/or evanescent
behaviour of the solutions and the local radial wave number. The
solutions are always evanescent or surface waves and the local
radial wave number is $k_z$. Note that in our notation
$\Gamma(\omega^2) = - k_z^2$.

\subsection{Incompressible MHD waves on uniform flux tubes}
For a uniform plasma we can rewrite (\ref{PInC1}) as
\begin{equation}
\frac{ d^2 P'}{ dr^2} +
\frac{ 1} { r} \frac{ d
P'}{ dr} - \left \{ \frac{
m^2}{ r^2} + k_z^2 \right \}  P' = 0. \label{PInC2}
\end{equation}
Equation~(\ref{PInC2}) can then be solved in terms of Bessel
functions $I_m(x)$ ($x = k_z r$) in the internal part of the flux
tube and $K_m(x)$ in the exterior region.
\begin{eqnarray}
 P'_{\mathrm i}(r) & = & \alpha I_m(x), \nonumber \\
\xi_{r, \mathrm{i}}(r) & = & \alpha \frac{ k_z}{
\rho_{\mathrm i} (\omega^2 - \omega_{\mathrm {Ai}}^2)}
\;I'_m(x), \nonumber \\
 P'_{\mathrm e}(r) & = & \beta K_m(x), \nonumber \\
\xi_{r, \mathrm{e}}(r) & = & \beta \frac{ k_z} {
\rho_{\mathrm e} (\omega^2 - \omega_{\mathrm {Ae}}^2)} \;  K'_m(x).
\label{SolutionsInC1}
\end{eqnarray}
Continuity of total pressure
and the radial component of the Lagrangian displacement leads to
the dispersion relation:
\begin{equation}
F \;\;\frac{ I'_m(x_0) K_m(x_0)}{
I_m(x_0) K'_m(x_0)} = 1. \label{DRInC1}
\end{equation}
The incompressible version of $F$ is
\begin{equation}
F = \frac{ \rho_{\mathrm e} (\omega^2 - \omega_{\mathrm {Ae}}^2)}
{ \rho_{\mathrm i} (\omega^2 - \omega_{\mathrm {Ai}}^2)}. \label{FInC}
\end{equation}
Now $x_0 = k_z R$. The dispersion relation (\ref{DRInC1}) can be
solved numerically. However, if we consider the TT approximation, 
the Bessel functions $I_m(x)$ and $K_m(x)$  in (\ref{DRInC1}) are
replaced with their first order asymptotic expansions, and the
dispersion relation (\ref{DRInC1}) is reduced to
\begin{eqnarray}
1 + F = 0.
\label{DRInC2}
\end{eqnarray}
The solution is again given by equation~(\ref{Freqkink}), i.e. $\omega^2  = \omega_{\mathrm k}^2$.
The important point to note is that we get exactly the same
expression for the frequency in the incompressible limit as in the
pressureless limit. As far as the frequency is concerned the MHD
waves with $m \geq 1$ do not make any distinction between a
pressureless plasma and an incompressible plasma at least in the
thin tube approximation. Here also  $\Lambda$ is given by
(\ref{RatioGradPrT2}) $\Lambda_{\mathrm i} (\omega^2) = - \Lambda_{\mathrm i} (\omega^2)
=  \frac{ \rho_{\mathrm i} - \rho_{\mathrm e}} { \rho_{\mathrm i} +
\rho_{\mathrm e}}$. This shows that the MHD waves are always dominated by
magnetic tension forces and that they are predominantly Alfv\'{e}nic
in nature. Since the fast waves are absent as we have imposed the
condition of incompressibility, the magnetic pressure force is
associated with the slow wave behaviour of the solution. The
eigenfunctions are
\begin{eqnarray}
\frac{ \xi_{r, \mathrm{i}}(r)}{ R} & = & C, \nonumber \\
\frac{ \xi_{\varphi, \mathrm{i}}(r)}{ R} & = & i \;C, \nonumber \\
\frac{ \xi_{z, \mathrm{i}}(r)}{ R} & = & i C (k_z R)
\frac{ r} { R},\nonumber \\
\frac{ P'_{\mathrm i}(r)}{ (B^2/\mu)} & = & C \;(k_z R)^2 \;
\frac{ \rho_{\mathrm i} - \rho_{\mathrm e}}{ \rho_{\mathrm i} + \rho_{\mathrm e}} \;
\frac{ r}{ R},\nonumber \\
\nabla \cdot \vec{\xi}_{\mathrm i} & = & 0, \nonumber \\
\xi_{\varphi, \mathrm{e}}(R_{>}) & = & - \xi_{\varphi, \mathrm{i}}(R_{<}),\nonumber \\
\xi_{z, \mathrm{e}}(R_{>}) & = & - \xi_{z, \mathrm{i}}(R_{<}).
\label{SolutionsInC2}
\end{eqnarray}
As before we have omitted terms of order $(k_z R)^2$ and higher
unless the terms of order $(k_z R)^2$ are the first non-vanishing
contribution to the expression under study.
It is instructive to compare (\ref{SolutionsInC2}) with
(\ref{SolutionsNoPr2}). The expressions for $\xi_{r, \mathrm{i}}(r)/ R$,
$\xi_{\varphi, \mathrm{i}}(r)/ R$ and $P'_{\mathrm i}(r)/(B^2/\mu)$ are exactly the same
as in the pressureless case. There is now a small (of the order
$(k_z R)$ axial component $\xi_z$ of the Lagrangian displacement.
This axial component now makes $\nabla \cdot \vec{\xi}$ exactly
equal to zero while it is of order $(k_z R)^2$ for a pressureless
plasma. The azimuthal component $\xi_{\varphi, \mathrm{i}}(r)$ and the axial
component $\xi_{z, \mathrm{i}}(r)$ are $\pi/2$ out of phase respect to
$\xi_{r, \mathrm{i}}(r)$. $\xi_{r, \mathrm{i}}(r)$ and $\xi_{\varphi, \mathrm{i}}(r)$ have equal
magnitude and are constant in the flux tube. The axial component
$\xi_{z, \mathrm{i}}(r)$ is of the order $(k_z R)$ and varies linearly in the
loop. In the thin tube approximation $\xi_{z, \mathrm{i}}(r)$ is always small
compared to $\xi_{r, \mathrm{i}}(r)$ and $\xi_{\varphi, \mathrm{i}}(r)$.  The dominant
motion is in horizontal planes normal to the equilibrium magnetic
field. The vertical displacement and velocity are small, of the order of $k_z$ at
the tube boundary. As in the
pressureless case $\xi_{\varphi}$  is discontinuous at $r=R$ with
opposite values at $R_<$ and $R_>$. This discontinuous behaviour is
due to the change of sign of the factor $\omega^2 - \omega_{\mathrm A}^2$. In
the incompressible limit also $\xi_{z}$ is discontinuous at $r=R$
with opposite values at $R_<$ and $R_>$. This discontinuous behaviour
is due to the change of sign of the factor $\omega^2 - \omega_{\mathrm C}^2$
with $\omega_{\mathrm C}^2 = \omega_{\mathrm A}^2$ in the incompressible limit.

\subsection{Beyond the TT approximation for incompressible MHD waves
on uniform flux tubes}

As in the compressible case the analytic expressions~(\ref{SolutionsInC2}) have
been obtained in the limit $k_z R << 1$. It is easy to solve the dispersion
relation (\ref{DRInC1}) and determine the spatial
solutions~(\ref{SolutionsInC1}). In Figure~\ref{eigen} the eigenfunctions of
three loops with variable radii are represented with circles. Again it is clear
that the spatial profile is well described by the approximated solutions in the
TT limit given by equations~(\ref{SolutionsInC2}). As expected, the case with
$R/L=0.1$ shows the largest deviation from the TT approximation.
Figure~\ref{eigen} also allows us a direct comparison with the eigenfunctions in
the compressible approximation. As the ratio $R/L$ decreases the eigenfunctions
of the compressible and incompressible cases tend to be the same, in agreement
with the analytical expressions given by (\ref{SolutionsInC2}) and
(\ref{SolutionsNoPr2}). We arrive to the same conclusion for the forces, even for
thick loops, the tension dominates over the magnetic pressure gradient.

\subsection{Incompressible MHD waves on non-uniform flux tubes}

We again remove the discontinuous variation of density from its internal value
$\rho_{\mathrm i}$ to $\rho_{\mathrm e}$ by a continuous variation in a non-uniform. By doing so,
we allow interaction of the global kink wave with local Alfv\'{e}n/slow
continuum waves and the
discontinuous behaviour of $\xi_{\varphi}$ and $\xi_{z}$ are replaced
by singular behaviour in ideal MHD and by large but finite values in
non-ideal MHD. In the thin boundary approximation we need to add an
additional term to the dispersion relation which takes into account the jump in
the radial component across the resonant layer where the real part of the kink
eigenmode is equal to the local Alfv\'{e}n frequency.  For the incompressible
case the jump in $\xi_r$ is \citep[see e.g.][]{Goossens1992}
\begin{eqnarray}
[\xi_r] &=& -i \pi \frac{ m^2 / r_{\mathrm A}^2 + k_z^2} {
\rho \mid \Delta_{\mathrm A} \mid } \;\;P',\nonumber \\
\;\; [P']&=& 0.
\label{JumpInC}
\end{eqnarray}
In the incompressible limit $\omega_{\mathrm A} = \omega_{\mathrm C}$ and the
Alfv\'{e}n resonance and slow resonance coincide.  The jump in
$\xi_r$ due to the Alfv\'{e}n resonance is given in
(\ref{JumpNoPr}). The jump in $\xi_r$ due to the slow resonance is
\begin{eqnarray}
[\xi_r] &=& -i \pi \frac{  k_z^2}{ \rho \mid \Delta_{\mathrm C} \mid}
\left \{ \frac{ v_{\mathrm S}^2}{ v_{\mathrm S}^2 + v_{\mathrm A}^2} \right \}^2 P',\nonumber \\ 
\;\; [P']&=& 0 ,\;\;
\Delta_{\mathrm C} = \frac{d}{dr}\left(\omega^2 - \omega_{\mathrm C}^2\right).
\label{JumpSlR}
\end{eqnarray}
For the incompressible limit  $v_{\mathrm S}^2 \rightarrow \infty$
(\ref{JumpSlR}) becomes
\begin{eqnarray}
[\xi_r] &=& -i \pi \frac{  k_z^2}{ \rho \mid \Delta_{\mathrm C} \mid}
 P',\nonumber \\
\;\; [P']&=& 0 , \;\;  \Delta_{\mathrm C} = \Delta_{\mathrm A} = \frac{d}{dr}\left(\omega^2 -
\omega_{\mathrm A}^2\right).
\label{JumpSlRInC}
\end{eqnarray}
We see that equation~(\ref{JumpInC}) is the combination of the jump due to the Alfv\'{e}n
resonance (\ref{JumpNoPr}) and the jump due to the slow resonance
(\ref{JumpSlRInC}). The contribution to the jump  in $\xi_r$ due to
the slow resonance is of order $(k_z R)^2$ compared to that of the
Alfv\'{e}n resonance and can be neglected in our thin tube
approximation. Note that $(v_{\mathrm S}^2/(v_{\mathrm S}^2 + v_{\mathrm A}^2))^2 \leq 1$ so that
the slow resonance has its biggest effect for incompressible
plasmas. Even in that case it is unimportant compared to the
Alfv\'{e}n resonance. This is in agreement with a  result obtained
for MHD waves in prominences by \cite{Soler2009}.  For sentimental
reasons we shall keep the contribution due to the slow resonance.

The modified version of the ideal dispersion relation (\ref{DRInC1})
is
\begin{equation}
F \;\;\frac{ I'_m(x_0) K_m(x_0)}{
I_m(x_0) K'_m(x_0)} -i \; G \; \frac{
K_m(x_0)}{  K'_m(x_0)}
= 1. \nonumber \\
\label{DRInC3}
\end{equation}
$F$ is given by (\ref{F}) and $G$ is defined as
\begin{equation}
G = \pi \frac{ m^2 / r_{\mathrm A}^2  + k_z^2}{\rho \mid \Delta_{\mathrm A} \mid }
\frac{
\rho_{\mathrm e} \left(\omega^2 - \omega_{\mathrm {Ae}}^2\right)}{ k_{\mathrm e}}.
\label{GInC}
\end{equation}
When we combine the TT approximation with TB approximation, the dispersion
relation is reduced to
\begin{equation}
1 + F  - i \; G \; \frac{ k_z R}{ m} = 0.
\label{DRInC4}
\end{equation}
The zero order solution to (\ref{DRInC4}) without taking into account
the effect of the resonance is of course (\ref{Freqkink}). The effect
of the resonance is contained in $G$. In order to take that effect
into account we proceed as before
\begin{equation}
\omega = \omega_{\mathrm R} + i \gamma, \;\;\omega_{\mathrm R} = \omega_{\mathrm k},
\label{OmegaComplex}
\end{equation}
and approximate $\omega^2$ with $\omega_{\mathrm k} ^2 + 2 i \omega_{\mathrm k} \gamma$.
The solution for the damping decrement is
\begin{equation}
\frac{ \gamma}{  \omega_{\mathrm k}} = -
\frac{ \pi/2}{
\omega_{\mathrm k}^2} \left \{\frac{ m}{ R} +
\frac{ (k_z R)^2}
{ m R} \right \} \frac{ \rho_{\mathrm i}^2 \rho_{\mathrm e}^2}
{ (\rho_{\mathrm i} + \rho_{\mathrm e})^3}
\frac{ \left(\omega_{\mathrm {Ai}}^2 - \omega_{\mathrm {Ae}}^2\right)^2}{\rho(r_{\mathrm A}) \mid
\Delta_{\mathrm A}(r_{\mathrm A}) \mid}.
\label{GammaIC1}
\end{equation}
If the variation of $\omega_{\mathrm A}^2$ is solely due to the variation of
density $\rho$ as is the case here since the vertical magnetic
field is constant, the equation can be rewritten as
\begin{equation}
\frac{ \gamma}{  \omega_{\mathrm k}} = -
\frac{ \pi}{ 8}
\left \{\frac{ m}{ R} +
\frac{ (k_z R)^2}{ m R}
\right \}\frac{ \left(\rho_{\mathrm i} - \rho_{\mathrm e}\right)^2}
{ \rho_{\mathrm i} + \rho_{\mathrm e} } \frac{ 1}
{\mid \frac{ d \rho}{ dr} \mid_{r_{\mathrm A}}}.
\label{GammaIC2}
\end{equation}
From here on we shall specialise to $m=1$. For a linear profile of
density
\begin{eqnarray}
\frac{ \gamma}{  \omega_{\mathrm k}} &=&
 - \frac{ \pi}{ 8}
\left (1 + (k_z R)^2 \right )
\frac{ l}{ R}\frac{ \rho_{\mathrm i} - \rho_{\mathrm e}}
{ \rho_{\mathrm i} + \rho_{\mathrm e}}, \nonumber \\
\frac{ \tau_{\mathrm D}}{ T} &=& \frac{ 4}
{ \pi^2}
\frac{ 1}{ l/R} \frac{ \rho_{\mathrm i} + \rho_{\mathrm e}}
{ \rho_{\mathrm i} - \rho_{\mathrm e}}
\left (1 - (k_z R)^2 \right).
 \label{GammaIC3}
\end{eqnarray}
For a sinusoidal profile of density
\begin{eqnarray}
\frac{ \gamma}{  \omega_{\mathrm k}} &=& - \frac{ 1}
{ 4}
\left (1 + (k_z R)^2 \right )
\frac{ l}{ R}\frac{ \rho_{\mathrm i} - \rho_{\mathrm e}}
{ \rho_{\mathrm i} + \rho_{\mathrm e}},\nonumber \\
\frac{ \tau_{\mathrm D}}{ T} &=& \frac{ 2}
{ \pi}
\frac{ 1}{ l/R} \frac{ \rho_{\mathrm i} + \rho_{\mathrm e}}
{
\rho_{\mathrm i} - \rho_{\mathrm e}}\left (1 - (k_z R)^2 \right).
 \label{GammaIC4}
\end{eqnarray}
For all practical purposes we can neglect the contribution
proportional to $(k_z R)^2$ and conclude that the damping due to
resonant absorption of the kink mode in an incompressible plasma
is the same as that in a pressureless plasma (see equations~(\ref{GammaNoPr3}) and
(\ref{GammaNoPr4})). If we forget about
differences proportional to $(k_z R)^2$  then the conclusion is
that kink MHD waves in pressureless plasmas and incompressible
plasmas are the same. In view of that conclusion it is difficult
to understand why a kink mode can be called fast as fast waves are
absent from incompressible plasmas.

The eigenfunctions in the thin dissipative layer can be described by
the functions $\tilde{F}(\tau)$ and $\tilde{G}(\tau)$ which were first
introduced by \cite{Ruderman1995} for non-stationary incompressible
resonant Alfv\'{e}n waves in planar plasmas. The conclusion is the same
as in the previous section. The  kink MHD waves are  highly Alfv\'{e}nic
in the dissipative layer.

\section{MHD kink waves in the presence of MHD radiation}

\subsection{Equations for compressible MHD waves on  a non-zero beta
plasma cylinder} 

So far we have seen that kink MHD waves in the thin
tube approximation do not care about propagating (body wave) or
evanescent (surface wave) behaviour in the internal part of the flux
tube. The behaviour in the exterior plasma was until now evanescent.
Here we take the next step and consider leakage of energy due to MHD
radiation. MHD radiation causes the frequencies to be complex even
in absence of resonant damping. MHD waves in the presence of MHD
radiation were studied for uniform flux tubes by \cite{Spruit1982}
in the TT approximation and by \cite{Cally1985, Cally2003} for
arbitrary values of the radius. \cite{Stenuit1998} and
\cite{Stenuit1999} determined MHD waves undergoing resonant
absorption and/or leakage for photospheric flux tubes embedded in a
non-magnetic surrounding. \cite{Stenuit1999} pointed out which Hankel
function to use for leaky and non-leaky waves.  We use the equations
for linear MHD waves on a 1-dimensional cylinder with  a straight
field. Effects due to plasma pressure and compressibility are taken
into account. The equations are
\begin{eqnarray}
D\frac{ d(r \xi_r)}{ dr}&  = &  - C_2 r P',
\nonumber \\
\rho \left(\omega^2 - \omega_{\mathrm A}^2\right) \xi_r & = &
\frac{ d P'}{ dr}, \nonumber \\
\rho \left(\omega^2 - \omega_{\mathrm A}^2\right) \xi_{\varphi} & = & \frac{i m}{r} P',
\nonumber \\
\rho \left(\omega^2 - \omega_{\mathrm C}^2\right) \xi_{z} & = & i k_{z}
\frac{ v_{\mathrm S}^2}{ v_{\mathrm S}^2 + v_{\mathrm A}^2} P',
 \nonumber \\
\nabla \cdot \vec{\xi} & = & \frac{ - \omega^2 \;
P'}{ \rho \left(v_{\mathrm S}^2 + v_{\mathrm A}^2\right) \left(\omega^2 -
\omega_{\mathrm C}^2\right)}.
\label{MHDwavesMHDR1}
\end{eqnarray}
The coefficient functions $D$ and $C_2$ are now
\begin{eqnarray}
D & = & \rho \left(v_{\mathrm S}^2 + v_{\mathrm A}^2\right) \left(\omega^2 -
\omega_{\mathrm A}^2\right)\left(\omega^2- \omega_{\mathrm C}^2\right), \nonumber \\
C_2 & = & \omega^4 -\left(v_{\mathrm S}^2 + v_{\mathrm A}^2\right)\left(\omega^2 -
\omega_{\mathrm C}^2\right)\left(\frac{m^2}{r^2} + k_z^2\right).
\label{MHDwavesMHDR2}
\end{eqnarray}
As before we rewrite the two first order differential equations
of (\ref{MHDwavesMHDR1}) as a second order ordinary differential
equation for $P'$:
\begin{equation}
\rho \left(\omega^2 - \omega_{\mathrm A}^2\right) \frac{ d }{ dr}
\left \{\frac{ r}
{ \rho \left(\omega^2 - \omega_{\mathrm A}^2\right)} \frac{ d P'}
{ dr} \right \} =
\left \{\frac{ m^2}{ r^2} - \Gamma(\omega^2)
 \right\} r P',
\label{PMHDR1}
\end{equation}
where $\Gamma(\omega^2)$ is now defined as
\begin{equation}
\Gamma(\omega^2) = \frac{\left(\omega^2 - k_z^2 v_{\mathrm S}^2\right)\left(\omega^2 -
\omega_{\mathrm A}^2\right)}
{\left(v_{\mathrm S}^2 + v_{\mathrm A}^2\right)\left(\omega^2 - \omega_{\mathrm C}^2\right)}.
\label{GammaMHDR}
\end{equation}
We have solved the set of equations (\ref{MHDwavesMHDR1}) under general
conditions allowing for non-zero plasma pressure and compressibility, see \citet
[]{Spruit1982,Cally1985} for uniform plasmas and \citet[]{Goossens1993} for
nonuniform plasmas. Here we present the results for a pressureless plasma with
$v_{\mathrm S}=0$. Equations (\ref{MHDwavesMHDR1}) are then reduced to equation
(\ref{MHDwavesNoPressure1}) and equation (\ref{GammaMHDR}) is reduced to
equation (\ref{Gamma}). On one hand, sine we aim to study MHD waves that are
propagating in the external medium we require that $\omega^2 - \omega_{\mathrm {Ae}}^2 >0$.
On the other hand, we want to show that kink waves do not need trapping and
since we want to have resonant absorption present in the model we require that
$\omega^2 - \omega_{\mathrm {Ai}}^2 <0$. In case of a constant magnetic field then the
inequalities imply that  $\rho_{\mathrm i} < \rho_{\mathrm e}$ so that the flux tube is underdense.
Note that when we allow $v_{\mathrm S} \neq 0$ and/or the equilibrium magnetic field to be
non-constant then MHD radiation does not necessarily require an underdense
loop.

\subsection{MHD waves on uniform flux tubes}

For a uniform plasma without gas pressure we recover equation (\ref{PNoPr2}).
Now $\Gamma_{\mathrm i}(\omega^2) < 0$ and $\Gamma_{\mathrm e}(\omega^2)>0$ and $\omega_{\mathrm {Ae}}^2<
\omega^2 <
\omega_{\mathrm {Ai}}^2$. The radial wave numbers are now defined as (see
equation (\ref{kike}))

\begin{equation}
k_{\mathrm i}^2 = -\Gamma_{\mathrm i}(\omega^2) = -\frac {\omega^2 - \omega_{\mathrm {Ai}}^2}
{ v_{\mathrm {Ai}}^2}, \;\;\; k_{\mathrm e}^2 = \Gamma_{\mathrm e}(\omega^2) = 
\frac {\omega^2 - \omega_{\mathrm {Ae}}^2}{ v_{\mathrm {Ae}}^2}.
\label{kikem}
\end{equation}

The solutions to equation (\ref{PNoPr2}) are now
\begin{eqnarray}
 P'_{\mathrm i}(r) & = & \alpha I_m(x), \nonumber \\
\xi_{r, \mathrm{i}}(r) & = & \alpha \frac{ 1}{ \rho
(\omega^2 - \omega_{\mathrm {Ai}}^2)} \; k_{\mathrm i} \;I'_m(x), \nonumber \\
 P'_{\mathrm e}(r) & = & \beta H^{(1)}_m(y), \nonumber \\
\xi_{r, \mathrm{e}}(r) & = & \beta \frac{ 1}{ \rho
(\omega^2 - \omega_{\mathrm {Ae}}^2)} \; k_{\mathrm i} H'^{(1)}_m(y).
\label{SolutionsMHDR1}
\end{eqnarray}
The formulation with the Hankel functions is convenient as it
enables us to distinguish between incoming and outgoing waves.
With the classic convention that $Re(\omega) >0$, and the time
dependency $\exp (-i \omega t)$, the function $H^{(1)}_m$ corresponds to an
outgoing wave and $H^{(2)}_m$ to an incoming wave. Hence we drop a
possible contribution due to $H^{(2)}_m$ since we are dealing with
the eigenvalue problem and not with the driven problem where the incoming wave is
prescribed. For the solutions interior in the tube we have taken
the Bessel function $I_m$ as in \cite{Goossens1992}.

Continuity of total pressure and the radial component of the
Lagrangian displacement leads to the dispersion relation:
\begin{equation}
F \; \frac{ I'_m(x_0) H^{(1)}_m(y_0)}{
I_m(x_0) H'^{(1)}_m(y_0)} = 1,
\label{DRMHDR1}
\end{equation}
where $F$ is defined by (\ref{F}). Now $x_0 = k_{\mathrm i} R$ and $y_0 = k_{\mathrm e} R$.
The dispersion relation (\ref{DRMHDR1}) can be solved numerically.
This was done by e.g. \cite{Spruit1982} and \cite{Cally1985,Cally2003}. Using the TT
 approximation the dispersion relation (\ref{DRMHDR1}) is reduced to
\begin{equation}
1 + F \;\frac{ k_{\mathrm e}}{ k_{\mathrm i}}\left \{1 + i \frac{\pi}{2}
(k_{\mathrm e} R)^2 \right \} = 0.
\label{DRMHDR2}
\end{equation}

When we neglect in a zeroth order approximation the effect of MHD radiation the
solution to equation (\ref{DRMHDR2}) is again equation (\ref{Freqkink}) for the
square of the frequency and equation (\ref{kekiNoPr}) for the radial wave
numbers.

We can rewrite equation (\ref{DRMHDR2}) correct up to second order in $(k_{\mathrm e} R)$ as
\begin{equation}
1 + F \;\frac{ k_{\mathrm e}}{ k_{\mathrm i}} = i \frac{\pi}{2} (k_{\mathrm e} R)^2
.\label{DRMHDR3}
\end{equation}
The solution is
\begin{eqnarray}
\omega_{\mathrm R} & = & \omega_{\mathrm k}, \nonumber \\
\frac{\gamma}{  \omega_{\mathrm k}} & = &
\frac{\pi} { 4} (k_{\mathrm e} R)^2
\frac{\rho_{\mathrm i} \rho_{\mathrm e}}{\left(\rho_{\mathrm i} +
\rho_{\mathrm e}\right)^2} \frac{\omega_{\mathrm {Ae}}^2 -
\omega_{\mathrm {Ai}}^2}{\omega_{\mathrm k}^2}. \label{GammaMHDR1}
\end{eqnarray}
Equation~(\ref{GammaMHDR1}) then takes the simple form
\begin{equation}
\frac{ \gamma}{  \omega_{\mathrm k}} =-
\frac{ \pi} { 8} (k_z R)^2
\frac{ \left(\rho_{\mathrm e} - \rho_{\mathrm i}\right)^2}{\left( \rho_{\mathrm i} + \rho_{\mathrm e}\right)^2}. \label{GammaMHDR2}
\end{equation}
The ratio of the force due to the pressure gradient to the magnetic tension force
is here also given by equation~(\ref{RatioGradPrTf}). 

The TT approximations to the eigenfunctions are again given by equations
(\ref{SolutionsNoPr2}). If we allow $v_{\mathrm S}\neq 0$ then we find a non-zero $\xi_{z}$ with 
\begin{eqnarray}
\frac{ \xi_{z, \mathrm{i}}(r)}{ R} \sim \frac{ v_{\mathrm {Si}}^2}
{ v_{\mathrm {Si}}^2 + v_{\mathrm {Ai}}^2} \; (k_z R) \; \frac{
r}{ R}.
\label{SolutionsNoPr4}
\end{eqnarray}

\subsection{MHD radiating  waves on non-uniform flux tubes}

Once more we remove the discontinuous variation of
density. Now the jump in
$\xi_r$ is given by (\ref{JumpNoPr}), and the modified version of the
ideal dispersion relation (\ref{DRMHDR1})  is
\begin{equation}
F \; \frac{ I'_m(x_0) H^{(1)}_m(y_0)}{
I_m(x_0) H'^{(1)}_m(y_0)}
- i G \frac{ H^{(1)}_m(y_0)}{  H'^{(1)}_m(y_0)}
= 1.
\label{DRMHDR4}
\end{equation}
The TT approximation to this equation (for $m=1$) is
\begin{equation}
1 + F \;\frac{ k_{\mathrm e}}{ k_{\mathrm i}}\left \{1 + i \frac{\pi}{2}
(k_{\mathrm e} R)^2 \right \} -
i G (k_{\mathrm e} R) \left \{1 + i \frac{\pi}{2} (k_{\mathrm e} R)^2 \right \} = 0.
\label{DRMHDR5}
\end{equation}
In the third term of the left hand side of the previous equation
we can drop $i \frac{\pi}{2} (k_{\mathrm e} R)^2$ for two reasons. First it
produces a term of order $(k_{\mathrm e} R)^3$, second it leads to a (small) change
in the real part of the frequency $\omega_{\mathrm R}$ which we approximate by $\omega_{\mathrm k}$. We can then rewrite (\ref{DRMHDR5})
correct up to second order in $(k_{\mathrm e} R)$ as
\begin{equation}
1 + F \;\frac{ k_{\mathrm e}}{ k_{\mathrm i}} = i \frac{\pi}{2}
(k_{\mathrm e} R)^2 + i G (k_{\mathrm e} R).
\label{DRMHDR6}
\end{equation}
The solution of (\ref{DRMHDR6}) is 
\begin{eqnarray}
\omega_{\mathrm R} & = & \omega_{\mathrm k}, \nonumber \\
\frac{ \gamma}{  \omega_{\mathrm k}} & = &
 -\frac{ \pi / 2}{ \omega_{\mathrm k}^2}
\frac{ 1}{ R}\frac{ \rho_{\mathrm i}^2
\rho_{\mathrm e}^2} { (\rho_{\mathrm i} + \rho_{\mathrm e})^3} \frac{
(\omega_{\mathrm {Ai}}^2 - \omega_{\mathrm {Ae}}^2)^2}{\rho(r_{\mathrm A}) \mid \Delta_{\mathrm A} \mid} \nonumber \\
&+&
\frac{ \pi}{ 4} (k_{\mathrm e} R)^2
\frac{ \rho_{\mathrm i} \rho_{\mathrm e}}{ (\rho_{\mathrm i} +
\rho_{\mathrm e})^2} \frac{ \omega_{\mathrm {Ae}}^2 -
\omega_{\mathrm {Ai}}^2}{  \omega_{\mathrm k}^2}. \label{GammaMHDR3}
\end{eqnarray}
In case of a constant magnetic field so that the variation of the
local Alfv\'{e}n frequency is solely due to a variation of density
(\ref{GammaMHDR3}) can be further simplified to

\begin{equation}
\frac{ \gamma}{  \omega_{\mathrm k}} = -
\frac{ \pi} { 8} \frac{ l}
{ R} \frac{ \rho_{\mathrm e} - \rho_{\mathrm i}}{
\rho_{\mathrm i} + \rho_{\mathrm e}} - \frac{ \pi} { 8} (k_z
R)^2 \frac{ \left(\rho_{\mathrm e} - \rho_{\mathrm i}\right)^2}{\left( \rho_{\mathrm i} +
\rho_{\mathrm e}\right)^2} < 0. \label{GammaMHDR4}
\end{equation}
Equation~(\ref{GammaMHDR4}) is derived for a linear variation of density. For a
sinusoidal variation the factor $\pi/8$ is to be replaced with $1/4$.  The
absolute value of the ratio of the decrement due to resonant absorption (first
term of equation~(\ref{GammaMHDR4})) and that due to MHD radiation (second
term of equation~(\ref{GammaMHDR4})) is
\begin{equation}
\frac{ l / R}{(k_z R)^2} \frac{\rho_{\mathrm e} + \rho_{\mathrm i}}{\rho_{\mathrm e} -
\rho_{\mathrm i}}.
\label{GammaMHDR5}
\end{equation}
This clearly indicates that damping due to resonant absorption dominates over
that due to MHD radiation (since $k_z R \ll 1$). For example, for a tube with $\rho_{\mathrm e}/\rho_{\mathrm i} = 3$ and 
$R/L=0.01$ then a very tiny layer of $l/R =5\times 10^ {-4}$ is enough for resonant
absorption to dominate over radiation.

The remarkable result is that the frequency of the kink wave and its damping due
to resonant absorption found for a compressible plasma with a non-zero plasma
pressure differ by terms of order $(k_z R)^2$, even when we allow MHD radiation.
If we neglect contributions proportional to $(k_z R)^2$ then the simple
conclusion is  that the frequency of the kink wave and its damping due to
resonant absorption are the same in the three cases that we have considered.

Again, as in the two previous cases, the eigenfunctions in the thin dissipative
layer can be described by the functions $\tilde{F}(\tau)$ and $\tilde{G}(\tau)$.
Again the conclusion is that kink MHD waves are highly Alfv\'{e}nic in the
dissipative layer.

\section{Conclusion}

This paper has examined the nature of MHD kink waves. This was done by
determining the frequency, the damping rate and in particular the eigenfunctions
of MHD kink waves for three widely different MHD waves cases: a compressible
pressureless plasma, an incompressible plasma and a compressible plasma with
non-zero plasma pressure which allows for MHD radiation. The overall conclusion
is that kink waves are very robust and do not care about the details of the MHD
wave environment. In all three cases the frequency and the damping rate are for
practical purposes the same as they differ at most by terms proportional to
$(k_z R)^2$. In the magnetic flux tube the kink waves are in all three cases, to
a high degree  of accuracy incompressible  waves with negligible pressure
perturbations and with mainly horizontal motions. The main restoring  force of
kink waves in the magnetised flux tube is the magnetic tension force.  The
gradient pressure force  cannot be neglected except when the frequency of  the
kink wave is equal or slightly differs from the local Alfv\'{e}n frequency, 
i.e. in the resonant layer. The adjective fast is not the correct adjective to
characterise  kink waves. If an adjective is to be used it should be
Alfv\'{e}nic. However, it  is better to realize that kink waves have mixed
properties and cannot be put in  one single box.

\begin{acknowledgements} This research was begun when M.G. was a visitor of the
Solar Physics Group at the UIB. It is pleasure for M.G. to acknowledge the warm
hospitality of the Solar Physics Group at the UIB  and the support received from
UIB. M.G. also acknowledges support from K.U.Leuven via GOA/2008-19 and J.T. from
GOA/2009-009. J.A. was supported by an International Outgoing Marie Curie
Fellowship within the 7th European Community Framework Programme. In addition,
J.T., I.A. and J.L.B. acknowledge the funding provided under projects AYA2006-07637
(Spanish Ministerio de Educaci\'on y Ciencia) and PCTIB2005GC3-03 (Conselleria
d'Economia, Hisenda i Innovaci\'o of the Government of the Balearic Islands).
\end{acknowledgements}

\end{document}